\documentclass[preprint,superscriptaddress,showpacs,preprintnumbers,amsmath,amssymb,aps,pre]{revtex4-1}

\usepackage[utf8]{inputenc} 
\usepackage{epsfig}
\usepackage{natbib}
\usepackage{amsmath}
\usepackage{subfigure}
\usepackage{epstopdf} 
\usepackage{graphicx}
\usepackage{color}
\usepackage{pict2e}
\newcommand{\comment}[1]{}

\DeclareMathOperator{\sinc}{sinc}

\begin{document}

\title{On the stability of uniform motion}

\author{\'Alvaro G. L\'{o}pez}
\email[]{alvaro.lopez@urjc.es}
\affiliation{Nonlinear Dynamics, Chaos and Complex Systems Group, Departamento de F\'{i}sica, Universidad Rey Juan Carlos, Tulip\'{a}n s/n, 28933 M\'{o}stoles, Madrid, Spain}

\date{\today}

\begin{abstract}
We show that the uniform motion of a homogeneous distribution of electric charge can be stable or unstable depending on its geometry. When the electrodynamic body is perturbed from a state of rest, it starts to perform fast oscillations, irrespective of the frequency of the perturbation. This nonlinear oscillation is the result of the feedback interaction between Coulombian and radiative fields. The resulting spontaneous symmetry breaking of the Lorentz group implies that the principle of inertia only holds on average and suggests that the default state of matter is not necessarily uniform motion, but self-oscillation as well. We propose that the excitability of  electrodynamic bodies under external perturbations, which leads to limit cycle oscillations, is at the basis of the wave particle duality and its related quantum effects.
\end{abstract}

\maketitle

\section{Introduction}

\begin{figure}[b]
	\centering
	\includegraphics[width=0.7\textwidth]{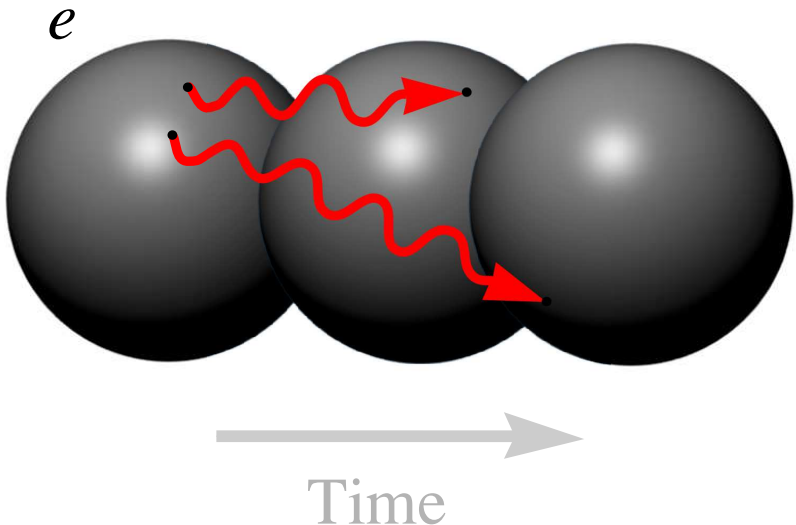}	\caption{A moving electron modeled as a spherical charge. At a certain instant of time the particle emits a field perturbation as a consequence of its accelerated motion, which travels in spacetime and affects other regions of the particle in the nearby future.}
	\label{fig:1}
\end{figure}

It has been recently demonstrated using a toy model of an extended moving prolate body that electromagnetic self-interactions can produce limit cycle oscillations \cite{lop20}. These interactions appear when such a body is accelerated and the radiation emitted by a certain region of the particle affects some other region of the particle at a later time (see Fig.~\ref{fig:1}). This phenomenon implies that charged sources experience self-forces and that the dynamics of extended bodies is fundamentally non-Markovian. Moreover, since the light cone evolves with the particle as it moves, the time delay depends on its kinematic status. The higher that it is the speed and the tangential acceleration of the body, the further from the past that come the self-signals affecting it at the present time.

The presence of self-forces on extended bodies can be mathematically described in terms of the retarded Li\'enard-Wiechert potentials \cite{lie98,wie01}, which are solutions to Maxwell's equations in the presence of sources \cite{max65}. In turn, these potentials lead to a description of the geodesic motion of the mass center of the body by means of state-dependent \emph{delayed differential equations} \cite{lop20}. As it is well known, most time delayed systems can harbour limit cycle behaviour through Hopf bifurcations \cite{lak11}. In addition to the traditional linear terms of inertia and harmonic oscillation, these functional differential equations introduce nonlinear terms of dissipation and antidamping as well. The resulting equation of motion exhibits the structure of the equation of a \emph{self-oscillator} \cite{jen13}. 

Self-oscillation can be described as the generation and the maintenance of an oscillatory motion by a source of energy that lacks any corresponding periodicity. Since there is no need of frequency tuning in self-oscillation, it is the system itself who controls with its dynamics the take and release of power from and to the environment. In this respect, self-oscillation is fundamentally different from resonance and relies on feedback and nonlinearity. In the case of electrodynamic bodies, the energy can come from the field ``of" the particle itself, which is distributed over the whole space. By taking energy from this field the particle can self-propel for a brief time when it is perturbed from a state of rest \cite{lop20}. However, at some point the different electromagnetic interactions do not act in phase anymore and the system decelerates, turning around towards its original location. This feedback mechanism is the result of an interplay between radiative fields and radiation reaction, the Abraham-Lorentz playing a crucial role in its development.
 
Importanly, as a consequence of the nonlinearity and the time delay, the electrodynamic of moving bodies becomes non-conservative. The dissipation and self-excitation of the body introduce an arrow of time in the dynamical system, insofar as the limit cycle can be run in one direction but not in the reverse. This fundamental irreversibility can be related to de Broglie's internal clock \cite{deb24} and its frequency is manifestly connected to the \emph{zitterbewegung} appearing in Dirac's equation. The resulting trembling motion of the particle explains the fact that a charged body can have a permanent electromagnetic wave attached to it. From this point of view, electrodynamic bodies should not be regarded as isolated systems, but might be better considered as open dissipative structures \cite{pri71}, which interact with their surrounding electromagnetic fields and operate far from equilibrium. Or, if prefered, and using the more modern jargon of complex systems theory, it can be said that particles are locally active \cite{chu05}. 

However, given the extremely prolate geometrical nature of the model used in previous works \cite{lop20}, it is not clear if there is a dependence on the geometry of the body for instability to take place. Simply put, it is hardly believable that the self-oscillations are always maintained irrespective of the nature of the perturbation and the geometry of the electrodynamic body. In what follows we use a continuous extended model of an electron to provide a rationale that allows to compute the stability of the particle's translational motion in terms of its geometry in any conceivable situation. We further demonstrate that the uniform motion of a spherical extended body is Lyapunov stable to translational perturbations. For this purpose, we perform a stability analysis of the rest state of the body by linearizing its equation of motion about the state of rest. Given the results of previous works \cite{lop20}, we conclude that the uniform motion of an electrodynamic body can be stable or unstable depending on its geometry and the nature of the perturbation. Therefore, we predict the existence of a \emph{Hopf bifurcation} as the shape of the body switches from a prolate geometry to an oblate one. We also suggest that, as a consequence of self-interactions, Newton's first and second laws only hold on average. We finally derive the frequency of vibration close to the fixed uniform trajectories and discuss the implications of our findings.

\section{Model description}

To model the electrodynamic body we use the Lagrangian density for classical electrodynamics, which is composed by two components: a field component and a source component. Note that in this representation sources of electric charge are given \emph{a priori}, assuming certain density of charge distributed in the space. As we discuss along the present work, this contrasts with a more fundamental model in which particles arise from the fields themselves, as some nonlinear electromagnetic waves, whose modes of vibrations describe the properties of the particle. Therefore, the Lagrangian density can be written as
\begin{equation}
\mathcal{L} =- \frac{1}{4 \mu_{0}} F_{\mu \nu}F^{\mu \nu}-A_{\mu}J^{\mu},
\label{eq:1}
\end{equation}
where $F^{\mu \nu}$ is the electromagnetic tensor and $J^{\mu}$ is the four current-density.

Then, the Maxwell's equations in covariant form can be derived from the previous action, which yields by differentiation the equations
\begin{equation}
\partial_{\mu} F^{\mu \nu}=\mu_{0} J^\nu,
\label{eq:1}
\end{equation}

In the present work we assume that the charged body has a spherical solid shape. We insist that further discussion on the nature of a fundamental particle is provided in the colophon of this letter in light of the present results. Therefore, the four-density of charge can be written as
\begin{equation}
J^\mu=\rho U^{\mu},
\label{eq:2}
\end{equation}
where $\rho$ is the volumetric density of charge in the proper frame and $U$ is the four-velocity. Since the body is assumed to maintain its shape and, in the present work, we are studying fluctuations in the non-relativistic limit, we can write the density of charge at time $t$ as $\rho(x,t)=\rho(x-x_{s}(t))$, where $x_{s}(t)$ represents the vector position of the mass centre of the corpuscle at time $t$. The solutions to Maxwell's equations are then expressed in terms of Jefimenko's equations \cite{jef92}. For the electric field we have
\begin{equation}
F^{i0}=\frac{1}{4 \pi \epsilon_{0}c} \int \left( \dfrac{R^i}{R^3}\rho+\dfrac{R^i}{c R^2}\partial_{t}\rho-\dfrac{1}{c^2 R}\partial_t J^i \right)_{t=t_r} d^3x',
\label{eq:3}
\end{equation}
while the magnetic field constitutes the purely spatial part of the tensor, which can be written as
\begin{equation}
F^{ij}=\frac{\mu_{0}}{4 \pi}\epsilon^{ijk}\epsilon_{klr} \int \left( \dfrac{R^{l}}{R^3}J^r+\dfrac{R^l}{c R^2}\partial_{t}J^r\right)_{t=t_r} d^3x',
\label{eq:4}
\end{equation}
where the retarded time $t_r=t-R/c$ has been introduced, with $R=|x-x'|$. In the present analysis, since we are linearizing these equations and the self-force contribution of the magnetic fields is of second order in the mass centre coordinates and its higher derivatives, we can also neglect Eq.~\eqref{eq:4}. 

\section{Stability analysis}

To perform the stability analysis of the electrodynamic body, we focus on Eq.~\eqref{eq:3} and write the self-force as
\begin{equation}
F_{\text{self}}^{i}=\int{c \rho F^{i 0}d^3x}.
\label{eq:5}
\end{equation}
Then, we express the charge density of the body by means of its Fourier transform as follows
\begin{equation}
\rho(x,t)=\dfrac{1}{(2 \pi)^{3}}\int \hat{\rho}(k)e^{-i k_j (x^j-x^j_{s}(t))} d^3k.
\label{eq:6}
\end{equation}
For a spherical charge it is straightforward to derive
\begin{equation}
\hat{\rho}(k)=3e\dfrac{\cos(k r)-\sinc(kr)}{(k r)^2},
\label{eq:7}
\end{equation}
where the constants $e$ and $r$ represent the charge and the classical radius of the electron, respectively.
Now we express the derivatives of the charge density and the vector current-density as
\begin{equation}
\partial_t \rho=-\partial_{i}\rho v^i_s(t),
\label{eq:8}
\end{equation}
and
\begin{equation}
\partial_t J^i=-\gamma v^i_{s}(t)\partial_{j}\rho v^j_s(t)+ \rho \partial_{t}\gamma v^i_{s}(t) +\rho \gamma a^i_{s}(t),
\label{eq:9}
\end{equation}
respectively. Since we aim at studying just perturbations of the linearized problem about the rest state, we can approximate the Eq.~\eqref{eq:9} by keeping only the third term in the right hand side hereafter. For the same reason, we neglect the Lorentz factor and we also consider that the gradient of $\rho$ appearing in Eq.~\eqref{eq:8}, which is evaluated at $x'-x_{s}(t)$, is simply evaluated at $x'$. At this stage, the self-force reads
\begin{equation}
F_{\text{self}}^{i}=\int \rho d^3x {\int { \left( \dfrac{\rho R^i}{R^3}-\dfrac{\partial_{j}\rho R^i}{c R^2} v^j_s(t)-\dfrac{\rho}{c^2 R} ~a^i_{s}(t) \right)_{t=t_r}} d^3x'},
\label{eq:10}
\end{equation}
where a factor $4\pi \epsilon_0$ has been omitted for aesthetic purposes. This constant factor will be reconsidered in the last expression of the self-force appearing in Eq.~\eqref{eq:15}. We now separately compute the three terms appearing in the right hand side of Eq.~\eqref{eq:10}. By using the Eqs.~\eqref{eq:6} and \eqref{eq:7} it can be shown, after some large algebraic manipulations, that the two first linearized terms appearing in the right hand side of Eq.~\eqref{eq:10} are simply zero. Therefore, we can approximate, to first order in the mass center coordinates $x_{s}(t)$ and their higher derivatives, the equation
\begin{equation}
F_{\text{self}}^{i}=-\int{\int{\dfrac{1}{c^2 R}\rho(x,t)\rho(x',t_r) a^i_{s}(t_r)} d^3x' d^3x}.
\label{eq:11}
\end{equation}
Moving to the dual space we obtain the self-force expressed in terms of the Fourier transform of the charge density as
\begin{equation}
F_{\text{self}}^{i}=-\dfrac{1}{(2 \pi)^3}\int{\hat{\rho}^2(k) d^3k \int{\dfrac{1}{c^2 R}e^{-i k_jR^j} a^i_{s}(t_r)}d^3R }.
\label{eq:12}
\end{equation}
To perform the stability analysis, we have to consider solutions of the form $x^i_{s}(t)=X^ie^{\lambda t}$, which are valid when the perturbations from the state of rest are small. In this case we obtain as second derivative $a^i_{s}(t)=X^i \lambda^2e^{\lambda t}$. Inserting this acceleration in the previous equation we derive
\begin{equation}
F_{\text{self}}^{i}=-\dfrac{\lambda^2 X^ie^{\lambda t}}{(2 \pi)^3c^2}\int{\hat{\rho}^2(k) d^3k \int{\dfrac{1}{ R}e^{-i k_jR^j} e^{-\lambda R/c}}d^3R }.
\label{eq:13}
\end{equation}
Integration over the whole configuration space by switching to spherical coordinates, then yields
\begin{equation}
F_{\text{self}}^{i}=-\dfrac{4 \pi \lambda^2 }{(2 \pi)^3} X^i e^{\lambda t} \int{ \dfrac{\hat{\rho}^2(k)}{k^2 c^2 +\lambda^2} d^3k}.
\label{eq:14}
\end{equation}
The solution to this integral leads to the self-force
\begin{equation}
F_{\text{self}}^{i}=-\dfrac{3 \hbar \alpha c}{2r^3} \left(\dfrac{c}{\lambda r} \right)^3  X^i e^{\lambda t} f(\lambda r/c),
\label{eq:15}
\end{equation}
where the characteristic polynomial $f(z) \equiv 2z^3-3z^2(1+e^{-2z})-6ze^{-2z}+3(1-e^{-2z})$ has been defined and the Plank's constant together with the fine structure constant have come into play with the help of equation
\begin{equation}
\dfrac{e^2}{4 \pi \epsilon_0 c}=\hbar \alpha,
\label{eq:16}
\end{equation}
which is formally known as Sommerfeld's relation \cite{som19}. At this point we recall that, as has been shown in previous works and can also be deduced from Eq.~\eqref{eq:10}, the force of inertia is already present in electromagnetism. More specifically, if the delay is neglected and we focus on the third term on the right hand side, such equation reads $F^i_{\text{self}}=-m a^i_s(t)$. The mass can be computed as the electrostatic energy of the sphere, which reads
\begin{equation}
m=\dfrac{1}{2 c^2}\int J_{\mu} A^{\mu} d^3x,
\end{equation}
\label{eq:17}
being $A^{\mu}$ the conventional electrostatic four-potential without retardation, and where Lorentz invariance has been used. We note that a factor of two has been introduced to avoid double counting, since a point of the body can never affect itself in the future, which would require the body to travel at speeds higher than light. In the case of a uniformly charged sphere it can be written as
\begin{equation}
m=\dfrac{3 \hbar \alpha}{5 c r},
\end{equation}
\label{eq:18}
where Sommerfeld's equation has been used again. This equation reveals that all sorts of physical momentum can be written as proportional to $\hbar$. As it has been shown if previous works, this equation and the electromagnetic origin of mass can be used to derive an analytical expression of the quantum potential \cite{lop20}. In brief, the lines above entail that mass has an electromagnetic origin and that Newton's second law must be written as a law of statics, following the principle of D'Alembert \cite{dal43}. If there are external forces present, we can write the equation of motion of an electrodynamic body as $F^i_{\text{self}}+F^i_{\text{ext}}=0$. If we insert Eq.~\eqref{eq:10} in this equation, the resulting equation of motion is a tremendously complicated functional differential equation, which manifests the mathematical nonlocal and nonlinear character of classical electromagnetism \cite{jac02}. 

In the present case, where there are no external forces, we simply have $F^i_{\text{self}}=0$. The solutions to this equation are the geodesic motions of the body. Therefore, we must compute the roots of $f(z)$ to find out the stability of the electrodynamic body.
\begin{figure}
	\centering
	\includegraphics[width=0.7\textwidth]{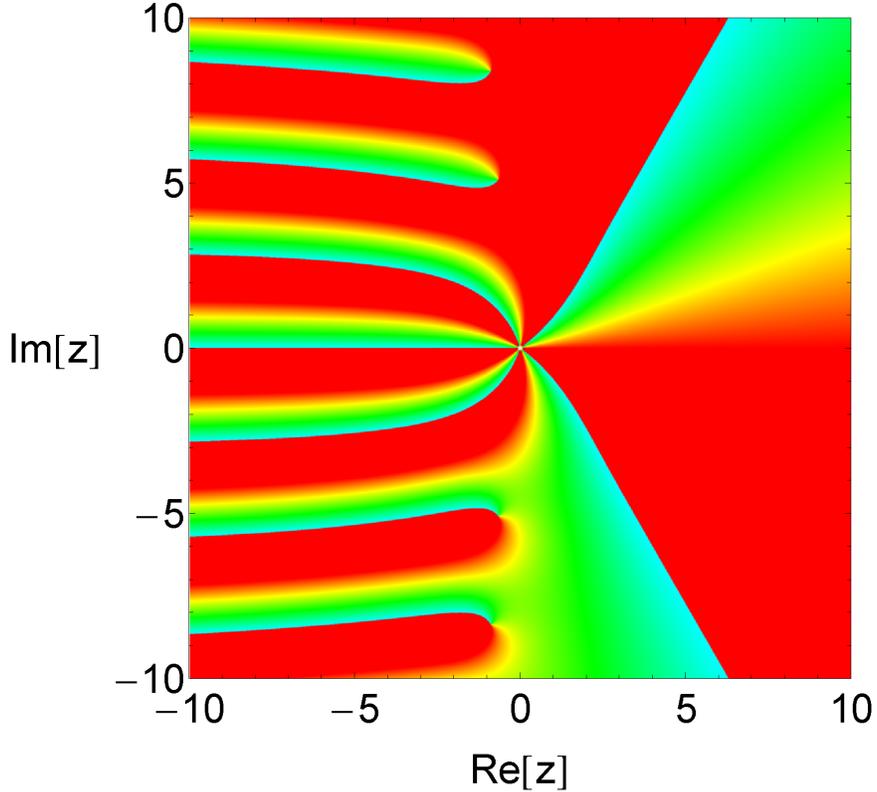}	
	\caption{A domain coloring representation of the function$f(z)=2z^3-3z^2(1+e^{-2z})-6ze^{-2z}+3(1-e^{-2z})$. The color represents the phase of the complex function. The roots can be localized where all the colors meet. In the present case we do not identify any roots with strictly positive real part. Therefore, the rest state is a stable fixed point for a spherical particle.}
	\label{fig:2}
\end{figure}
As can be seen in Fig.~\ref{fig:2}, there exist no roots with strictly positive real part for a spherical body, as suggested in previous works \cite{boh48}. This means that any brief harmonic pulse affecting the resting sphere, no matter how strong, will be damped away as times goes by. But importantly, we note that the decay is oscillatory, and that a pulse without periodicity can trigger a rapid transient oscillation with incommensurate frequency. Thus, even if an external force is not acting anymore on the particle, the particle would be oscillating, which contradicts the principle of inertia. Furthermore, any external sustained perturbation will produce a limit cycle self-oscillatory dynamics. This response is characteristic of self-oscillation, which does not require of a tuning between the frequency of an external source of energy and its internal vibration. From the eigenvalues it can also be seen that the coefficient of damping is proportional to $c/r$, just as it was the antidamping coefficient in previous works \cite{lop20}. Introducing the constants $\eta_{n}$, we also see that the frequencies of the several modes of oscillation can be estimated as
\begin{equation}
\omega_{n}=\eta_{n} \dfrac{c}{r},
\end{equation}
which reminds of the frequency of \emph{zitterbewegung} appearing in the solutions of the Dirac equation. Finally, we can operate a suitable change of reference frame to any other inertial observer by means of a Lorentz boost. Since Maxwell's equations are covariant, we can safely infer that any uniform motion of a spherical homogeneous distribution of charge is asymptotically stable in the Lyapunov sense. 
\begin{figure}
	\centering
	\includegraphics[width=0.7\textwidth]{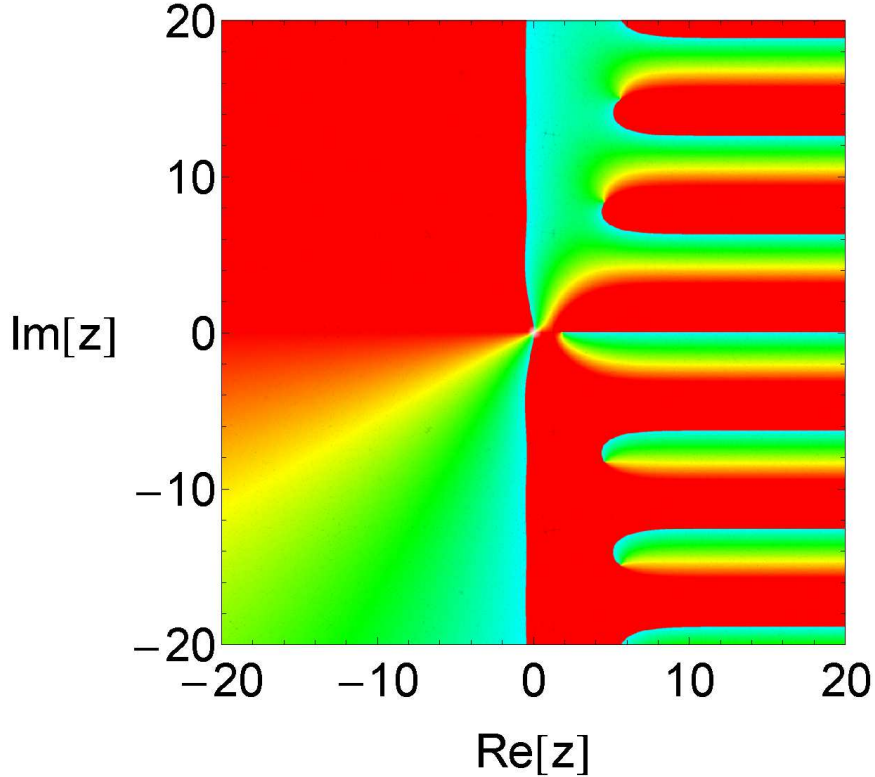}	
	\caption{The roots of the polynomial $f(z)=z^2+z+1-e^{z}$, which results from studying the stability of the charge distribution $\rho(x)=-e/2\delta(x^1)\delta(x^2)(\delta(x^3+r)+\delta(x^3-r))$, when a perturbation along an orthogonal axis $X^{\mu}=\delta^{\mu}_2$ is performed. This charge distribution consists of two point charges $e/2$ separated a fix distance $2r$. In this case all the roots present a strictly positive real part. Therefore, the rest state is a totally unstable fixed point for an extremely prolate distribution of charge.}
	\label{fig:3}
\end{figure}

\section{Discussion}

In summary, the present analysis constitutes further evidence suggesting that electrodynamic bodies can experience strong tidal self-forces as a consequence of radiation fields and radiation reaction. These time-delayed self-interactions can make them vibrate rather violently. However, as it has been shown, for a rotationless body with a spherical shape these fluctuations disappear as times goes by. This finding opposes to results found in previous works \cite{lop20}, in which a prolate geometry of the body was discovered to be unstable (see Fig.~\ref{fig:3}). Therefore, it is evident that the stability of an electrodynamic body depends on its geometry. Nevertheless, we note that even for a spherical shape, small external perturbations of any frequency can trigger transient fast oscillations. It is then expected that there exists a Hopf bifurcation as the shape of the particle changes from spherical to prolate, with respect its direction of motion. Likewise, this bifurcation can be produced by an external agent. Importantly, we recall that for a system with more than one particle, each member acquires several delays representing both self-interactions and mutual interactions as well. Then, these time delays can become coupled among them in their respective equations of motion, leading to the synchronization of self-oscillations and the subsequent entanglement of their dynamical states \cite{lop203}. 

In those situations in which the motion is unstable due to the prolate shape of the particle \cite{lop20}, these oscillations necessarily lead to the fact that such electrodynamic bodies have a pilot wave attached to it, as suggested originally by Louis de Broglie and also by recent experimental works \cite{cat08}. This internal fluctuation would explain in a simple and logical way typical phenomena that were believed to be exclusive of the mysterious quantum realm, as for example the double slit experiment or the capacity of fundamental particles to overcome external potential barriers. We recall that the possibility of a classical pilot-wave theory that shares properties with the quantum world has been recently confirmed by hydrodynamical experimental models of walking droplets \cite{for10}. 

The fact that Newton's second law can be derived from classical electromagnetism certainly argues in favor of electromagnetic mass. However, it is important to highlight the relevance of the two remaining terms appearing in Eq.~\eqref{eq:10} and also the fact that retardation introduces more terms when expanded in Taylor series. These terms are frequently disregarded in macroscopic physics since they are of higher order in $r/c$. However, in microscopic physics they can lead to the symmetry breaking of uniform motion and, therefore, to the violation of the principle of inertia \cite{lop20}. Moreover, as it has been shown in such work, the kinetic energy and an electrodynamic quantum potential can be derived from the Li\'enard-Wiechert potential by just assuming that mass is of electromagnetic origin. In this regard, Newton's first law should be perhaps considered an emergent law that holds on average for macroscopic bodies, but not necessarily in the microscopic realm. 

The proposal of a purely electromagnetic mass would have consequences of transcendental importance and deserves experimental attention. We cite a few logical consequences. In the first place, the gravitational field would have to be regarded necessarily as a weak residual electromagnetic field resulting from the trembling motions of protons, electrons and their mutual orbital motion. Noticeably, this key aspect would explain in very simple terms the principle of equivalence by identifying inertial and gravitational mass as plain electromagnetic mass. Furthermore, by just recalling that energy curves spacetime and including curvature effects, we would also see ourselves forced to admit that gravitational waves emerge from electromagnetic waves.

Concerning the nature of electrodynamic bodies, it is evident that rigid bodies must be disregarded as models of fundamental particles, since they are in contradiction with the principle of causality, which emanates from classical electromagnetism. But not less important, because they are unstable by themselves. Certainly, the charge distribution should repel itself leading to the explosion of the body. To solve this anomaly, Poincaré introduced artificial stresses to balance the repulsive forces \cite{poi05}.

Finally, if we assume that quantum fluctuations have an electromagnetic origin \cite{lop20}, we bear in mind that quantum field theories describe particles as vibrations of fundamental fields and recall that Einstein-Maxwell's equations are highly nonlinear partial differential equations, we are naturally led to think of fundamental particles as electromagnetic topological solitons \cite{ale80,fab01}. Although just a hypothesis, one is then inclined towards some sort of vortex rings \cite{kel67,kle13,ran95,arr17}, where the topological number could give the charge of the particle and the vorticity its spin. As a matter of fact, the present study points to the fact that electrons might not have spherical shape. Indeed, a vortex ring would have a prolate shape with respect to the direction of motion of its center of mass, what would guarantee the instability of uniform motion.

\end{document}